\documentclass[twocolumn,floatfix,superscriptaddress,amsmath,showpacs,showkeys,aps,pre]{revtex4-2}

\usepackage{environ,censor,calc}
\NewEnviron{mysout}
 {\soutrefunexp{\BODY}}
\newlength\nextcharwidth
\makeatletter
\renewcommand\@cenword[1]{%
  \setlength{\nextcharwidth}{\widthof{#1}}%
  \censorrule{\nextcharwidth}%
  \kern -\nextcharwidth%
  #1}
\makeatother

\usepackage{epsfig}
\usepackage{bm}
\usepackage[dvipsnames]{xcolor}
\usepackage{amsmath}
\usepackage{amssymb}
\usepackage{graphicx}
\usepackage{physics}
\usepackage{xr}
\usepackage{soul}
\usepackage{longtable}
\usepackage{verbatim}
\usepackage[OT1]{fontenc}
\usepackage[normalem]{ulem}
\usepackage{blkarray}

\begin{document}

\title{Data-driven inference of brain dynamical states from the $r$-spectrum of correlation matrices}
\author{Christopher Gabaldon}
 \affiliation{Instituto de Ciencias F\'isicas (ICIFI-CONICET), Center for Complex Systems and Brain Sciences (CEMSC3), Escuela de Ciencia y Tecnolog\'ia, Universidad Nacional de Gral. San Mart\'in, Campus Miguelete, San Mart\'in, Buenos Aires, Argentina}
  \affiliation{ {Departamento de Física, Facultad de Ciencias Exactas y Naturales, Universidad de Buenos Aires, Buenos Aires 1428, Argentina}
}
 \author{Adrià Mulero}
\affiliation{Facultat de Psicologia, Ciències de l'Educació i de l'Esport, Blanquerna, Universitat Ramon Llull,  Barcelona, Spain.}

\author{Rong Wang}
\affiliation{State Key Laboratory for Strength and Vibration of Mechanical Structures, School of Aerospace Engineering, Xi’an Jiaotong University, Xi’an 710049, China}

\author{Daniel A. Martin}
\affiliation{Instituto de Ciencias F\'isicas (ICIFI-CONICET), Center for Complex Systems and Brain Sciences (CEMSC3), Escuela de Ciencia y Tecnolog\'ia, Universidad Nacional de Gral. San Mart\'in, Campus Miguelete, San Mart\'in, Buenos Aires, Argentina}
\affiliation{Consejo Nacional de Investigaciones Cient\'{\i}fcas y T\'ecnicas (CONICET), Buenos Aires, Argentina}

\author{Sabrina Camargo}
\affiliation{Instituto de Ciencias F\'isicas (ICIFI-CONICET), Center for Complex Systems and Brain Sciences (CEMSC3), Escuela de Ciencia y Tecnolog\'ia, Universidad Nacional de Gral. San Mart\'in, Campus Miguelete, San Mart\'in, Buenos Aires, Argentina}
\affiliation{Consejo Nacional de Investigaciones Cient\'{\i}fcas y T\'ecnicas (CONICET), Buenos Aires, Argentina}

\author{Qian-Yuan Tang}
\affiliation{Department of Physics, Faculty of Science, Hong Kong Baptist University, Hong Kong SAR, China}

\author{Ignacio Cifre}
\affiliation{Facultat de Psicologia, Ciències de l'Educació i de l'Esport, Blanquerna, Universitat Ramon Llull,  Barcelona, Spain.}

\author{Changsong Zhou}
\affiliation{Department of Physics, Faculty of Science, Hong Kong Baptist University, Hong Kong SAR, China}

\author{Dante R. Chialvo}
\affiliation{Instituto de Ciencias F\'isicas (ICIFI-CONICET), Center for Complex Systems and Brain Sciences (CEMSC3), Escuela de Ciencia y Tecnolog\'ia, Universidad Nacional de Gral. San Mart\'in, Campus Miguelete, San Mart\'in, Buenos Aires, Argentina}
\affiliation{Consejo Nacional de Investigaciones Cient\'{\i}fcas y T\'ecnicas (CONICET), Buenos Aires, Argentina}
\affiliation{Department of Physics, Faculty of Science, Hong Kong Baptist University, Hong Kong SAR, China}

\date{\today}
\begin{abstract}
We present a data-driven framework to characterize large-scale brain dynamical states directly from correlation matrices at the single-subject level. By treating correlation thresholding as a percolation-like probe of connectivity, the approach tracks multiple cluster- and network-level observables and identifies a characteristic percolation threshold, $r_c$, at which these signatures converge. We use $r_c$ as an operational and physically interpretable descriptor of large-scale brain dynamical state. Applied to resting-state fMRI data from a large cohort of healthy individuals ($N=996$), the method yields stable, subject-specific estimates that covary systematically with established dynamical indicators such as temporal autocorrelations. Numerical simulations of a whole-brain model with a known critical regime further show that $r_c$ tracks changes in collective dynamics under controlled variations of excitability. By replacing arbitrary threshold selection with a criterion intrinsic to correlation structure, the $r$-spectra provides a physically grounded approach for comparing brain dynamical states across individuals.
\end{abstract}

\maketitle
The advent of magnetic resonance imaging, and in particular functional MRI (fMRI), has enabled the non-invasive investigation of large-scale brain dynamics by providing time-resolved measurements of distributed neural activity. This progress has motivated the widespread use of graph-theoretical descriptions, in which the brain is represented as a network whose nodes correspond to brain regions and whose edges encode functional or structural interactions. Within this framework, large-scale correlation matrices and their derived graphs are routinely compared across groups of subjects and experimental conditions to study how distributed neural circuits support cognition and behavior \cite{eguiluz2005,Sporns2010BCT,Sporns2010,MasudaReview}.

Despite its success, this approach commonly relies on fixed or heuristically chosen correlation thresholds. Thresholding compresses continuous correlation structure into a single discrete representation, reducing information about how connectivity emerges from collective interactions to a single scale and obscuring the balance between fragmented and coordinated configurations. This compression weakens the link between correlation structure and underlying brain dynamical state, particularly at the single-subject level, where a single connectivity snapshot cannot fully capture dynamical regime differences.

Here we address this limitation by reframing thresholding as a physically meaningful dimension of analysis rather than a technical nuisance. Instead of fixing a single threshold, we treat it as a control-like parameter whose variation probes collective changes in network organization. This perspective directly connects correlation-based brain networks to order--disorder transitions, in which tuning a control parameter reveals shifts between fragmented and globally coordinated collective states \cite{stanley,scale_invariance,Chialvo2010}. Within this framework, we introduce the \emph{$r$-spectrum}, which tracks multiple independent properties of the correlation network as a function of threshold. Their convergence at a characteristic value, denoted $r_c$, provides a physically grounded descriptor of large-scale brain dynamical state at the single-subject level. We then apply this framework to resting-state fMRI data and to a simple neuronal model to assess its empirical relevance and mechanistic interpretation.

\begin{figure*}[hb]
\centering
\includegraphics[width=0.9\linewidth]{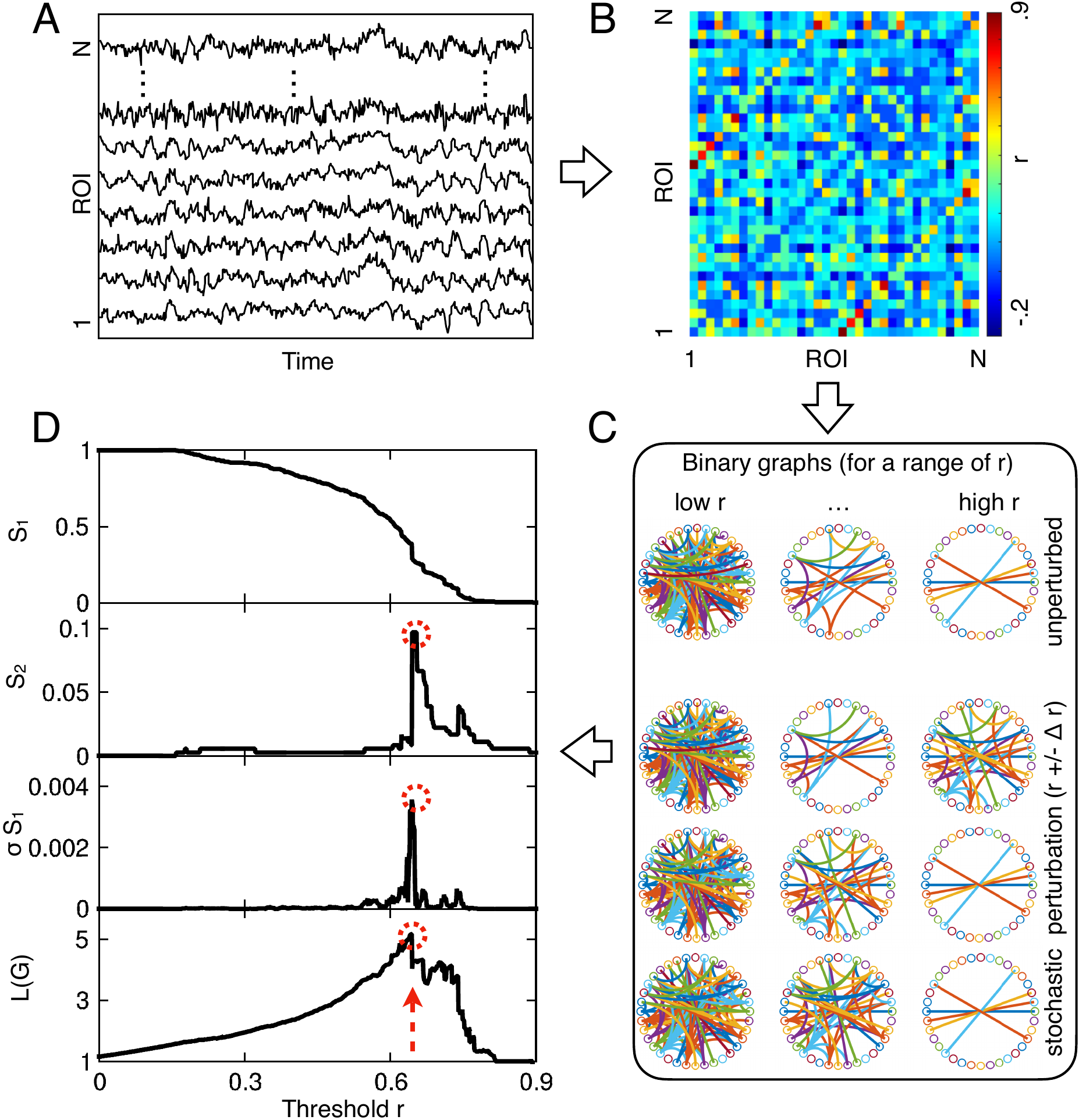}
 \caption{Typical workflow schematic to compute the $r$-spectra for a single subject brain data. From the BOLD time series in Panel A the correlation matrix is obtained (Panel B). The cartoon of Panel C shows the two types of binary graphs computed for a range of correlation thresholds $r$. One is calculated from the unperturbed thresholds $r$ (top graphs), and the others (bottom graphs) are stochastic realizations computed for small random variations around the thresholds ($r \mp \Delta r$). Panel D shows, for a typical subject, the ``$r$-spectra'': it plots, as a function of $r$,  the order parameter $S_1$ (top), and the three observables used to identify the graph percolation threshold, which correspond to the values of $r$ (arrow) for the maximum values (circles) of $S_2$,    $\sigma S_1$ and $L(G)$. Results from the first session of subject code \#100206 of the WU-Minn rs-fMRI dataset \cite{WuMinn}.}
    \label{fig:1}
\end{figure*}

\emph{Conceptual framework:} Fig.~\ref{fig:1} illustrates the workflow for constructing functional brain networks from resting-state fMRI data. Blood-oxygen-level-dependent (BOLD) time series are aggregated into regions of interest (ROIs; Fig.~\ref{fig:1}A), from which a correlation matrix is computed (Fig.~\ref{fig:1}B). An undirected binary graph is then obtained by connecting ROI pairs whose correlation exceeds a threshold $r$. While standard analyses compare network properties at fixed thresholds across groups, this work exploits the full threshold dependence of the correlation matrix. Rather than selecting a single $r$, we analyze how network structure varies with threshold and identify, for each subject, a characteristic percolation point separating fragmented configurations from ordered, system-spanning connectivity. As illustrated in Fig.~\ref{fig:1}C, high $r$ values yield sparse or disconnected networks, whereas decreasing $r$ leads to the emergence of a giant connected component, providing an operational marker of large-scale dynamical organization.

%%%%%%%%%%%%%%%%%%%%%%%%FIG1

At least three key observables (see also additional approaches \cite{Radicchi} and alternative definitions based on degree variability and cluster-size distributions in the Supplementary Material, SM) can be used to identify the critical point, hereafter denoted $r_c$ (Fig.~\ref{fig:1}D). First, within percolation theory, the transition separating disordered and ordered phases is expected to coincide with the maximum of the second-largest cluster size $S_2$. Second, the variability of the order parameter reaches a maximum at the transition, reflecting maximal competition between ordered and disordered configurations. In finite systems, $S_1$ serves as a proxy for the percolation order parameter, and its susceptibility, quantified by the variance under small perturbations, peaks at the transition. In practice, this is approximated by the peak of the variance of the largest cluster size, $\sigma S_1$, obtained under small perturbations of the control parameter $r$.
Third, from graph theory, the critical point is marked by a peak in the network characteristic path length $L(G)$. Together, these three observables, dubbed the ``$r$-spectra'', consistently identify the same characteristic threshold $r_c$, reflecting the concurrent appearance of multiple independent signatures associated with critical points in percolation theory. 
Although the networks considered here are finite and modular, rather than infinite lattices where percolation theory is exact, finite-size scaling theory predicts that these signatures remain well defined, with peak locations providing reliable estimates of the transition point. The connection between $r_c$ and standard percolation theory is further explored in the SM, where a clear anticorrelation between $r_c$ and analytical estimates of the critical bond density $p_c$ is observed.

\emph{Experimental results:} We analyzed human fMRI resting state and diffusion tensor imaging (DTI) recordings from 996 healthy subjects (ages 22-36)  belonging to  WU-Minn HCP Young Adult dataset \cite{WuMinn}.  For rs-fMRI data, the recording on each subject was repeated four times using high resolution 3T scanners, with time samples  $TR=0.72s$, for 1200 time steps.

As illustrated in Fig.~\ref{fig:1} for a typical single subject, the three ``$r$-spectrum'' observables exhibit maxima at approximately the same threshold $r$. Typically, $S_2$ peaks at $\sim$10--20\% of system size, coinciding with a rapid increase in $S_1$ (both normalized by system size: $0 \leq S_2 \leq S_1 \leq 1$). The variability $\sigma S_1$, computed over many realizations with stochastic perturbations of $r$, peaks where $S_1$ has its steepest increase, which coincides with the $S_2$ maximum. These cluster-size changes are reflected in graph metrics such as the characteristic path length $L(G)$, which also reaches a maximum at the percolation threshold.

\begin{figure}[hb!]
\centering
\includegraphics[width=	1.\linewidth]{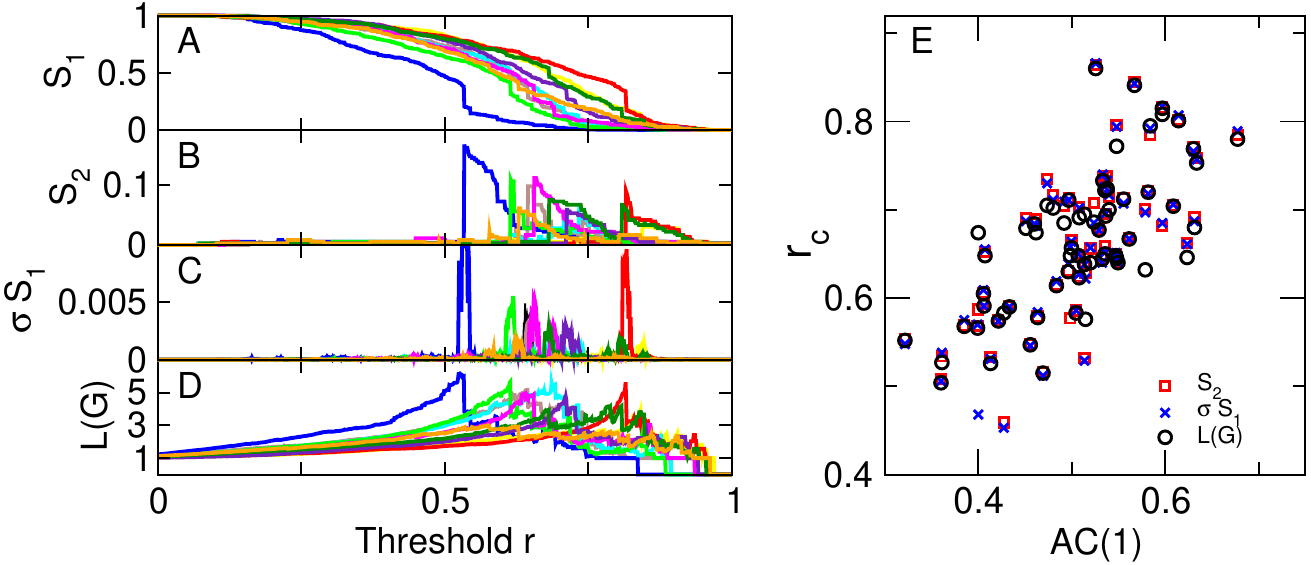}
 \caption{Across subjects $r_c$ covaries with the temporal correlation  of the average brain BOLD activity $AC(1)$. Panels A-D show, for a subset of ten typical subjects, the order parameter $S_1$ (top),  $S_2, \sigma S_1$ and $L(G)$ as a function of $r$. Panel E shows for 64 subjects the value of $AC(1)$ versus the values of $r_c$  derived from the maximum values of the observables $S_2, \sigma S_1$ and $L(G)$. }
    \label{fig:2}
\end{figure}

These results support our central hypothesis that $r_c$ is a measure of the brain's dynamic degree of order/disorder, in the precise sense that it marks the threshold separating fragmented (disordered) and system-spanning (ordered) correlation structures, averaged over time.
However, for this to hold, $r_c$ must correlate also  with established benchmarks of neural order/disorder, such as segregation-integration \cite{PNAS_Wang,WangPRL} or temporal and spatial correlations \cite{camargo,grigera}. Thus, we challenged this conjecture on a large repository of resting-state brain imaging data previously studied \cite{PNAS_Wang}. The results strongly support this hypothesis: as shown in Fig. \ref{fig:2}, $r_c$ covaries  with the first autocorrelation coefficient, $AC(1)$ \cite{Control}, of the mean BOLD signal. In individuals with low/high $r_c$, the brain's mean field activity shows markedly weaker/stronger autocorrelations (Fig. \ref{fig:2}E). 
We remark that temporal correlations quantified by $AC(1)$ are a classical manifestation of critical slowing down, and they have been repeatedly shown to be robust markers of proximity to a critical point in both models and empirical data \cite{AC_as_marker,AC_as_marker1,AC_as_marker2,Control,tagliaPropofol}.

\begin{figure}[ht!]
\centering
\includegraphics[width=	1.\linewidth]{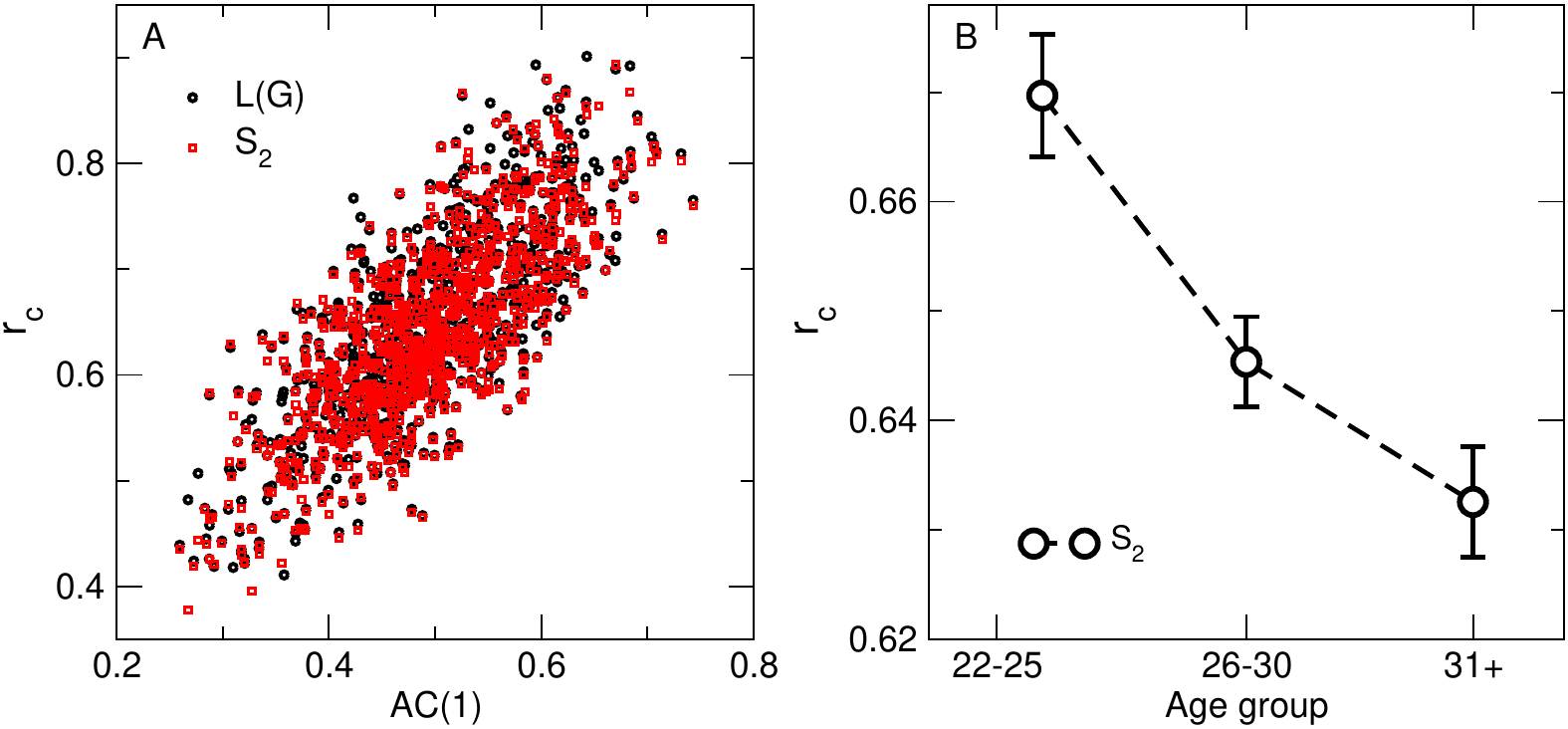}
 \caption{Panel A: The brain dynamical states  evaluated by the percolation thresholds $r_c$ strongly correlate in each of 996 individuals with the $AC(1)$ values computed from the global brain BOLD activity.   
Panel B: On average, the percolation threshold $r_c$ decreases with age. Age groups composition: 22-25 years (218 subjects), 26-30 years (429 subjects) and  31+ years 
 (349 subjects, 340 subjects of age 31-35 and 9 subjects of age 36+). Bars represent within-group standard errors.} % (i.e., within-group standard deviation divided by the square root of the number of subjects in that group minus one).}

 \label{fig:3}
\end{figure}

This relationship is further confirmed across the full dataset ($N=996$), revealing a clear population-wide trend in Fig.~\ref{fig:3}A. Estimates of $r_c$ obtained from graph-theoretical ($L(G)$) and percolation-based ($S_2$) measures are highly consistent. Results based on $\sigma S_1$ show the same behavior and are omitted for clarity; see SM for results across all four recording sessions. Consistent with percolation theory, inter-subject variability in $S_1$ peaks at $r_c$ (see SM). Grouping subjects by age further reveals a systematic decrease of $r_c$ with age (panel B), indicating weakened inter-areal correlations, consistent with reduced functional connectivity in aging (discussed below).

%%%%%%%%%%%%%%%%%Figure 2 %%%%%%%%%%%%%%%%%

%%%%%%%%%%%%%%%%%%%FIG3%%%%%%%%%%%%%%%%%%
%%%%%%%%%%%%%%%%%%%%%%%%%%%%%%%%%%%%%%%%

\emph{Model results:} We next contrast results from human brain data with those from a simple dynamical model \cite{Haimovici} implemented on an empirical neuroanatomical connectome \cite{Hagmann08}. This model, previously shown to reproduce key features of spontaneous brain activity, consists of a coarse-grained structural network of brain regions and dynamical rules governing their activity. The connectivity matrix in the Haimovici model \cite{Haimovici} is derived from neuroanatomical connectivity, describing the average fiber tract density between pairs of brain areas as estimated from diffusion-weighted and diffusion tensor imaging (DWI/DTI). For the simulations presented here, we use the normalized average DTI connectivity matrix obtained from $N=996$ subjects in the HCP dataset (see \emph{Methods} for details).

For simplicity, the dynamics of each node follow discrete-state excitable dynamics based on the Greenberg–Hastings model \cite{Greenberg}. Thus, each node is assigned one of three states: quiescent $Q$, excited $E$, or refractory $R$,  and the transition rules are: 1) $Q \rightarrow E$  with a small probability $p_1$ ($ \sim 10 ^{-3}$), or if the sum of the connection weights $w_{ij}$ with the active neighbors ($j$) is higher than a threshold $T$, i.e., $\sum w_{ij} > T$ and $Q  \rightarrow Q$  otherwise; 2) $E \rightarrow R$  always; 3) $R \rightarrow Q$ with a small probability $p_2$ ($\sim 10 ^{-1}$)  delaying the transition from the $R$ to the $Q$ state for some time steps. We fixed $p_1$ and $p_2$, which set the time scales of self-excitation and recovery, and varied $T$, the control parameter of the node dynamics (see \emph{Methods}). Note that $T$ is an intrinsic excitability threshold and is distinct from the correlation threshold $r$ used to construct functional networks.

%%%%%%%%% Figure 4%%%%%%%%%%%%%%%%%%
\begin{figure}[ht!]
\centering
\includegraphics[width=	.6\linewidth]{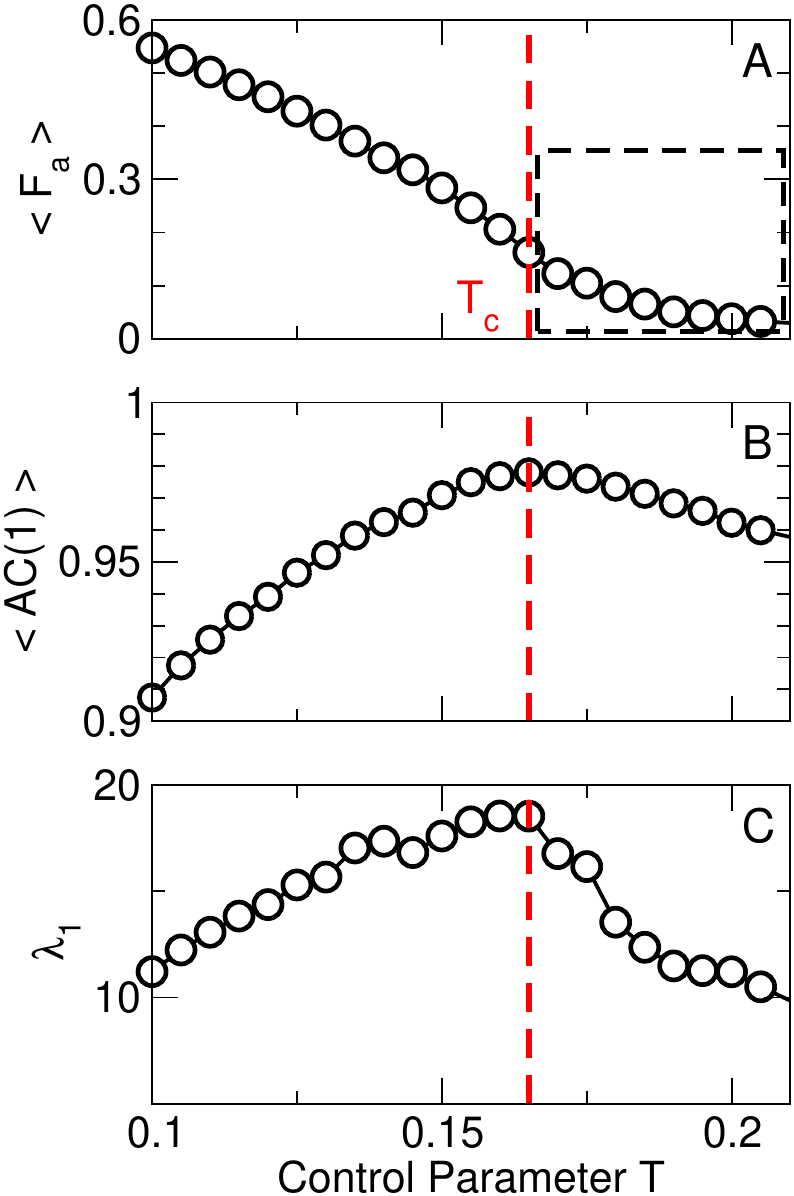}
\caption{Identification of the model' critical point $T_c$ (vertical dashed line) by the maximum in the autocorrelation of the order parameter fluctuations as well as the largest eigenvalue ($\lambda _1$) of the correlation matrix. Panels A-B show the average activation fraction $F_a$ serving as the order parameter and $AC(1)$  as a function of the control parameter $T$. The box in A denotes the range of $T$  explored in the results of  Fig. \ref{fig:5}.}
 \label{fig:4}
\end{figure}

To compare with the experimental results, we  simulate  various degrees of proximity to criticality, mimicking a variety of dynamical conditions which hypothetically are equivalent to those observed in the human subjects. For that we first run the model  for a relatively wide range of  $T$ and locate precisely the critical point $T_c$ (simulations lasted $10.000$ steps for each $T$ value, discarding the transient first 1000 steps). We track the $AC(1)$ of the time series of the fraction of active nodes ($F_a$)  as a function of $T$ and identify its maximum (vertical red dashed line in Fig. \ref{fig:4}). Then we run the model for several values of $T$ near $T_c$ collecting for each one relatively long time series of the model' nodes activity, and proceed to compute the respective pair wise correlations matrices. These matrices are comparable to those collected from the human brains fMRI. Finally, following the same workflow used for the human brains  (Fig. \ref{fig:1}C and D), we proceed to construct the binary graphs and calculate the $r_c$ values for each of the conditions. We expect to find that $r_c$ covaries with the model' $T$ under which the matrices were generated as well as with the correlation properties of the  order parameter (i.e.,  $F_a$).

\begin{figure}[ht!]
\centering
\includegraphics[width=	.8\linewidth]{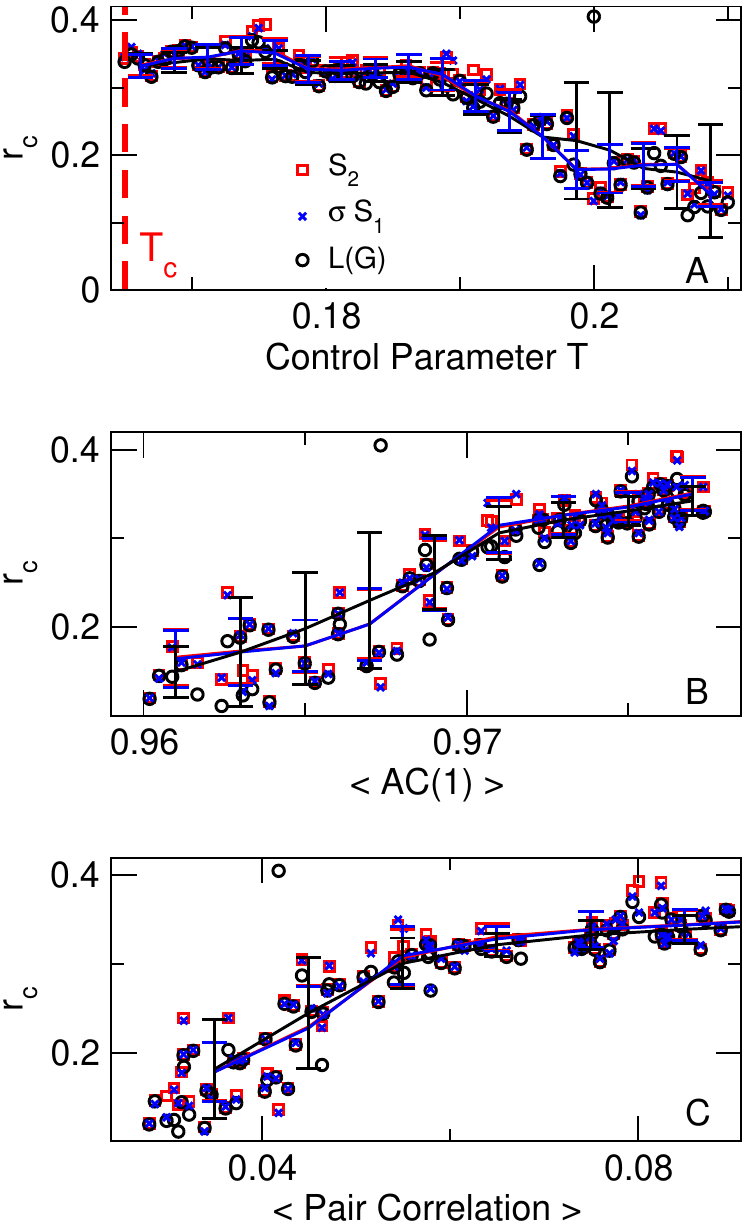}
\caption{The model dynamical state can be successfully estimated from the percolation thresholds $r_c$ (panel A), which in turn strongly correlates with the $AC(1)$ (panel B) and pair wise correlation (panel C) values computed from the time series of the mean field activity $F_a$. The red dashed vertical line in panel A indicates the value of the model critical point $T_c$. Continuous lines show mean $\pm$ standard deviation for  all points on a window of of size $\Delta T=0.0025$ (Panel A), $\Delta AC(1)=0.002$ (Panel B) or  $\Delta C=0.01$ (Panel C).} %Five times more simulation runs were used to compute standard deviation curves.} }
 \label{fig:5}
\end{figure}

The results presented in Fig.~\ref{fig:5} show a remarkable agreement with the hypothesis: different degrees of proximity to criticality are consistently accompanied by a monotonic shift in $r_c$ computed from the correlation matrices (see Fig.~\ref{fig:5}A). Each symbol corresponds to a single simulation run under a specific dynamical condition, that is, a given value of $T$. These results can be directly compared with Figs.~\ref{fig:2}E and \ref{fig:3}A, where each symbol represents a single human subject. We emphasize that the model is used as a proof of principle to provide a mechanistic interpretation of the observed trends, rather than to quantitatively match empirical values.
 
The analysis of the model' results reproduces the experimental observations in fMRI data (shown in Figs. \ref{fig:2}E and \ref{fig:3}A) demonstrating that $r_c$ covaries with $AC(1)$ (Fig.\ref{fig:5}B). For completeness, we  illustrate also a related correlation index, the average pair wise correlation, which is known to covary with $AC(1)$ (see \cite{refAC_CC} and SM) and consequently with  $r_c$  as seen in Fig.\ref{fig:5}C.
%%%%%%%%% Figure 5%%%%%%%%%

\emph{Discussion:} 
 This work shows that principles from percolation theory provide a natural framework for characterizing individual differences in large-scale brain dynamics. Even within apparently homogeneous groups, individual subjects exhibit distinct dynamical set points of functional connectivity, variability that has often been ignored or averaged out despite being noted in prior studies \cite{Makse1,Makse2,PNAS_Wang,WangPRL}. The $r$-spectrum approach addresses this by identifying, for each subject, a characteristic percolation threshold $r_c$ at which multiple independent signatures converge. The physiological interpretation is direct: $r_c$ estimates proximity to a percolation transition associated with efficient information exchange across scales. In conventional functional connectivity terms, lower or higher $r_c$ values correspond to weaker or stronger average connectivity, respectively (Fig.~5C). Analysis of resting-state network participation reveals that cortical networks display maximal fluctuations in cluster membership near $r_c$, whereas subcortical regions remain comparatively stable (see SM).

 Beyond simple correspondence with average connectivity, the $r$-spectrum framework provides convergent observables ($S_2$, $\sigma S_1$, $L(G)$) that jointly characterize dynamical state and help distinguish genuine organization from thresholding artifacts. The strong covariation between $r_c$ and temporal autocorrelations $AC(1)$ (Fig.~3A) reflects an underlying dynamical mechanism: critical slowing down increases temporal persistence of activity fluctuations, which in turn enhances spatial correlations across brain regions over finite observation windows. This temporal-to-spatial translation manifests as elevated $r_c$, providing a structural signature of proximity to criticality. Importantly, this relationship emerges from the concurrent behavior of multiple independent measures rather than a single statistic, indicating collective reorganization of correlation structure.

 The changes in $r_c$ observed across subjects closely mirror those obtained in numerical simulations under different levels of excitability (Fig.~5), supporting a common mechanistic interpretation: shifts in excitability move the system relative to a dynamical critical point, thereby altering correlation structure and the associated percolation threshold. Structural connectivity differences may also contribute to this variability. Consistent with this interpretation, the negative correlation between $r_c$ and age indicates a gradual shift toward more subcritical dynamics across the lifespan, in agreement with recent work using graph neural networks to characterize neurodevelopmental differences \cite{AHDH}.

 The $r$-spectrum framework also connects naturally to spectral measures of brain organization. In a recent study of the same dataset \cite{IntSeg}, subjects with \emph{Integrated} activity (characterized by a dominant leading eigenvalue $\lambda_1$) showed higher General Cognitive Ability, while those with \emph{Segregated} activity (greater weight in higher-order eigenvalues) showed higher Crystallized Intelligence and Processing Speed. Within the present framework, larger $\lambda_1$ corresponds to stronger global correlations and higher $r_c$, consistent with enhanced large-scale coordination. Conversely, stronger segregation is associated with weaker global correlations and lower $r_c$. Thus, $r_c$ provides a complementary descriptor linking spectral measures of integration--segregation to dynamical organization near criticality. While the present analysis focuses on resting-state fMRI data, the $r$-spectrum framework is in principle applicable to any system exhibiting critical behavior. This generality is illustrated in the SM through applications to the two-dimensional Ising model and the Greenberg--Hastings model on synthetic Watts--Strogatz \cite{WS} networks, which yield consistent results. Additional analyses of cluster-size distributions at $r_c$ are also presented in the SM.

In summary, we introduce a data-driven framework to infer large-scale brain dynamical state from correlation structure alone, grounded in principles from percolation and graph theory. Rather than fixing an arbitrary correlation threshold, the approach exploits threshold dependence to extract a characteristic scale $r_c$ that serves as an operational descriptor of dynamical regime, reflecting both proximity to optimal large-scale coordination and subject-specific organizational features.

%\newpage
 
{\emph{\bf Methods:}}

\emph{Datasets and data processing:} The fMRI data processing steps  follow exactly those in a previous study \cite{PNAS_Wang}, including  the resting state fMRI (rs-fMRI) and diffusion tensor imaging (DTI) recordings from  996 healthy subjects (ages 22-36) 
belonging to  WU-Minn HCP dataset.  For rs-fMRI data, each subject was registered 4 times using high resolution 3T scanners, with time windows  $TR=0.72s$, for 1200 time steps (864s for each registration) \cite{WuMinn}. We consider each recording separately.

\emph{Functional connectivity:}
We used the MRI data preprocessed by the HCP through the Minimal Processing Pipeline (MPP), which performs several processing steps, including spatial artifact/distortion removal, surface generation, cross-modal registration, and alignment to standard space \cite{Glasser2013}. 
 
The preprocessed resting-state fMRI data were converted to surface space (``CIFTI” format), which consists of 91282 cortical and subcortical gray matter coordinates with a resolution of 2 mm \cite{WuMinn}. 
The global whole-brain signal was not removed in this analysis \cite{Scholvinck}.
 
The brain  was parcellated into 360 regions according to the multimodal parcellation (MMP) atlas \cite{Glasser2016}  so that blood-oxygen level dependent (BOLD) time series for each region could be obtained by averaging voxel signals located within regions. Data was  band-pass  filtered on the range 0.01-0.1Hz \cite{Shine,Sobczak,Whittaker,WangPRL}.  Preprocessing was completed separately for each session.

\emph{Structural Connectivity:} The matrices of structural connectivity extracted from DTI were used for  numerical simulations of brain activity. Probabilistic tractography was performed on the DTI data to trace the white matter fibers bridging brain regions with the  default settings in the FSL software (version 5.0.9, http://www.fmrib.ox.ac.uk/fsl/) \cite{Glasser2016, Liu2020, Liu2020b}. 

First, the 360 MMP regions defined in the standard space were projected into individual cortical surfaces in the diffusion space by using the Connectome Workbench \cite{WuMinn}. Probabilistic tractography was then performed among pairs of brain regions with the FSL software \cite{Behrens, Behrens2}. During pair-wise fiber tracing, one region was defined as the seed area to start the tracing and the other as the target, and vice versa. Each voxel of the seed region will send out a total of 5000 streamlines. Streamline propagation were performed with a step length of 0.5 mm and a maximum step of 2000. 
After fiber tracing, the directional connective probability $p_{ij}$ from a seed region i to a target region j can be calculated  as the ratio between the number of streamlines that reached the target region and the total number of streamlines initiated from the seed region 
 \cite{Behrens}. Based on the calculated directional connective probability, we further defined the structural connectivity weight $w_{ij}$,
 by calculating the reciprocal averages of the connective probabilities (i.e. $w_{ij} = (p_{ij} + p_{ji})/2$).

\emph{Numerical simulation details:} We computed numerical simulations of the Haimovici model  on a network described by the structural connectivity matrix of the human subjects \cite{Haimovici, Zarepour, Sanchez,rocha1,rocha2}. 
Given the well known subsampling of the inter-hemispheric tracks the numerical simulation reported in the main text were conducted on the right  hemisphere, consisting on 180 ROIS (see SM for similar results on the left hemisphere). 
GH model is a simple cellular automaton model where each element $i=1,..,180$ represents a region or a group of neurons which is able to transition between 3 states: quiescent ($S_i(t) = Q$), excited ($S_i(t) = E$) or refractory ($S_i(t) = R$). At time $t + 1$ a quiescent node can become active due to an external input with a small probability $p_1$, or if the contribution of all active connections at time $t$ is larger than a threshold $T$ ($\sum_j w_{ij} \delta_{S_j(t),E} > T$, where $w_{ij}$ is the SC matrix with $i$,$j$ running from 1 to 180, and $\delta_{i,j}=1$ if $i=j$ and $0$ otherwise);  an active node will became refractory always, and a refractory node will become quiescent with probability $p_2$ (following \cite{Haimovici, Zarepour,Sanchez} we have used $p_1 = 10^{-3}$ and  $p_2 = 0.2$ throughout the text).  For the purposes of this work the network excitability (i.e., the interaction strength between nodes) is described by  the normalized structural connectivity matrix  $ w_{ij} \to w_{ij}/\sum_{j} w_{ij}  $ \cite{rocha1,rocha2}. The activation fraction is defined as $F_a(t)=\sum_i \delta_{S_i(t),E}$.

Previous work \cite{Haimovici, Zarepour, Sanchez,Control,rocha1,rocha2} shows that the model exhibits a subcritical regime with low activity for high values of $T$, a supercritical regime with high activity at low values of $T$, and a critical regime for intermediate values of $T$.  The critical regime can be readily identified from the value of $T$ that maximizes the first autocorrelation coefficient, $AC(1)$  of the activation fraction timeseries, $F_a(t)$. \cite{Sanchez, Control}  (See Fig. \ref{fig:4}).  Results for the same model, for arbitrarily large Watts-Strogatz matrices with connection weights mimicking experimental distributions are shown in SM.

\emph{Detailed description of the workflow to compute the $r$-spectra:} The $r$-spectra describes the changes  of the correlation matrix properties as a function of a cutoff parameter $r$. The aim is to identify relatively abrupt changes which may be straightforward indicatives of a phase transition in the network connectivity. To compute it, for each subject session or for each numerical simulation run, we analyzed time series consisting of 360 or 180 signals  lasting 1200 or $10^5$-$10^6$ samples respectively. For each session/run, we computed the Functional Connectivity (FC) matrix as the Pearson correlation coefficients of each pair of signals. Then, we proceed to binarize the FC matrix, specifically, for each value of $r$ in the range $0<r<1$ (with steps $\Delta r= 0.001$), we computed the adjacency matrix $M_{ij}^{r}$, which is 1 if $FC_{ij}>r$ and 0 otherwise. After that, we extracted the size and identities of all clusters from each matrix $M^{r}$ using  the reverse Cuthill-McKee algorithm \cite{reverse} implemented by the \emph{adj2cluster.m}  Matlab routine \cite{adj2clust}. We extracted the size of the largest $S_1$ and the second largest $S_2$ cluster for each session/run. In addition, for each $r$ value we computed the characteristic path length $L(G)$, defined as the average shortest-path distance between nodes in the largest cluster. In bond percolation, this behavior is well understood: when $p \ll p_c$, clusters are small and path lengths are short. As $p \to p_c^{-}$, clusters become large and ramified, with very long shortest paths even between spatially proximate nodes. At $p_c$, an infinite cluster first appears and the system becomes scale-invariant (fractal), with path lengths diverging in the thermodynamic limit. While brain graphs differ from bond percolation on lattices, the characteristic behavior of $L(G)$ near the transition remains qualitatively similar.

To identify the threshold $r_c$ we use three approaches (see alternatives in SM):  In the first one, for each session/run we extract  $r_c$ as the value of $r$ that maximizes $S_2$.  In the second one, for each session/run we computed several realizations with each matrix $M^{r}$  adding in each one a small stochastic perturbations ($\sim 10^{-3})$ of the control parameter $r$ and computed the variance of the largest cluster's size $\sigma S_1$. For this approach  $r_c$ is defined by the $r$ value that correspond to the largest $\sigma S_1$. 
In the third approach, from graph theory, $r_c$ is marked by a peak in the  largest cluster average path length $L(G)$.  For each session/run we computed for each value of $r$ the mean path length of the graph using routines written in Matlab based on the functions \emph{graph.m} and \emph{distance.m}.   

\emph{Data and Code Availability:} fMRI, DTI, and behavioral measures can  be accessed from the  WU-Minn HCP  https://balsa.wustl.edu/. The codes used in this study are available in GitHub at https://github.com/DanielAlejandroMartin/r-spectra

 \section*{Acknowledgements}
DRC thanks Hong Kong Baptist University for funding his Distinguished Professorship of Science during these studies. 
This research was supported by Natural Science Foundation of China (Nos. 12305052, and 12272292), Research Grants Council of Hong Kong (Nos. 22302723, 12202124, and SRFS2324-2S05), Guangdong and Hong Kong Universities “1+1+1” Joint Research Collaboration Scheme (2025A0505000011), and Hong Kong Baptist University's funding support (RC-FNRA-IG/22-23/SCI/03, and RC-SFCRG/23-24/SCI/06).

%%%%%%%%%%%%%%%%%%%%%%%%%%%%%%%%%%%%%%%%%%%%%%%%%%%%%%%%%%%
%%%%%%%%%%%%%%%%%%%%%%%%%%%%%%%%%%%%%%%%%%%%%%%%%%%%%%%%%%%
%%%%%%%%%%%%%%%%%%%%%%%%%%%%%%%%%%%%%%%%%%%%%%%%%%%%%%%%%%%
%%%%%%%%%%%%%%%%%%%%%%%%%%%%%%%%%%%%%%%%%%%%%%%%%%%%%%%%%%%
%%%%%%%%%%%%%%%%%%%%%%%%%%%%%%%%%%%%%%%%%%%%%%%%%%%%%%%%%%%
%
%
%                   SUPP MATERIAL
%
%
%%%%%%%%%%%%%%%%%%%%%%%%%%%%%%%%%%%%%%%%%%%%%%%%%%%%%%%%%%%
%%%%%%%%%%%%%%%%%%%%%%%%%%%%%%%%%%%%%%%%%%%%%%%%%%%%%%%%%%%
%%%%%%%%%%%%%%%%%%%%%%%%%%%%%%%%%%%%%%%%%%%%%%%%%%%%%%%%%%%
%%%%%%%%%%%%%%%%%%%%%%%%%%%%%%%%%%%%%%%%%%%%%%%%%%%%%%%%%%%
%\newpage
\clearpage
\onecolumngrid
%\appendix
%\newcommand{\mm}[1]{\langle #1 \rangle}
%%%%%%%%%%%%%%%%%%%%%%%%%%%%%%%%%%%%%%%%%%%%%%%%%%%%%%%%%%%%%%%%%%%%%%%%%%%%%%

\renewcommand{\thefigure}{S\arabic{figure}}
\renewcommand{\thesection}{S\arabic{section}}
\setcounter{figure}{0}    
\setcounter{page}{1}

\begin{center}
{\large \bf Supplemental Material  for\\ Data-driven inference of brain dynamical states from the $r$-spectrum of correlation matrices\\ by Gabaldon \emph{et al.}}
\end{center}
\vspace{1cm}

We include Supplemental data supporting the main manuscript. First, in Section  \ref{Extra} we reproduce results from Figures 2 and 3 of main text considering all sessions of each subject from HCP dataset. %We also show results for $r_c$ estimated from $\sigma S1$, omitted in Fig 3 of main text for clarity. 
We also include numerical simulation results of GH model on the left hemisphere structural network, analogous to those presented in Fig. 5 of main text for the right hemisphere.
In Section \ref{general} we apply the $r$-spectra method on two simple numerical models: the 2D Ising model describing paramagnetic-ferromagnetic transitions, and the GH model (the same as in main text), now using synthetic (Watts-Strogatz \cite{smWS}) networks. We also propose alternative definitions of $r_c$. In Section \ref{percolation}, we show further  insights on the relation among $r$-spectra and percolation. We show  the relation among $r_c$ and the estimated critical density on bonds, $p_c$. We also show the variability maximization of the largest cluster size about $r_c$, as expected about percolation.  In Sec. \ref{correlations} we relate temporal and spatial (Pearson) correlations under simple assumptions (for vector autorregressive models), as mentioned in main text. Finally in Sec. \ref{localizaition} we study the behavior of the resting state networks about $r_c$.

\section{Complementary figures}\label{Extra}

\subsection{$r$-spectra results for all sessions in the HCP dataset} 

\begin{figure}[hb!]
\centering
\includegraphics[width=0.6\linewidth]{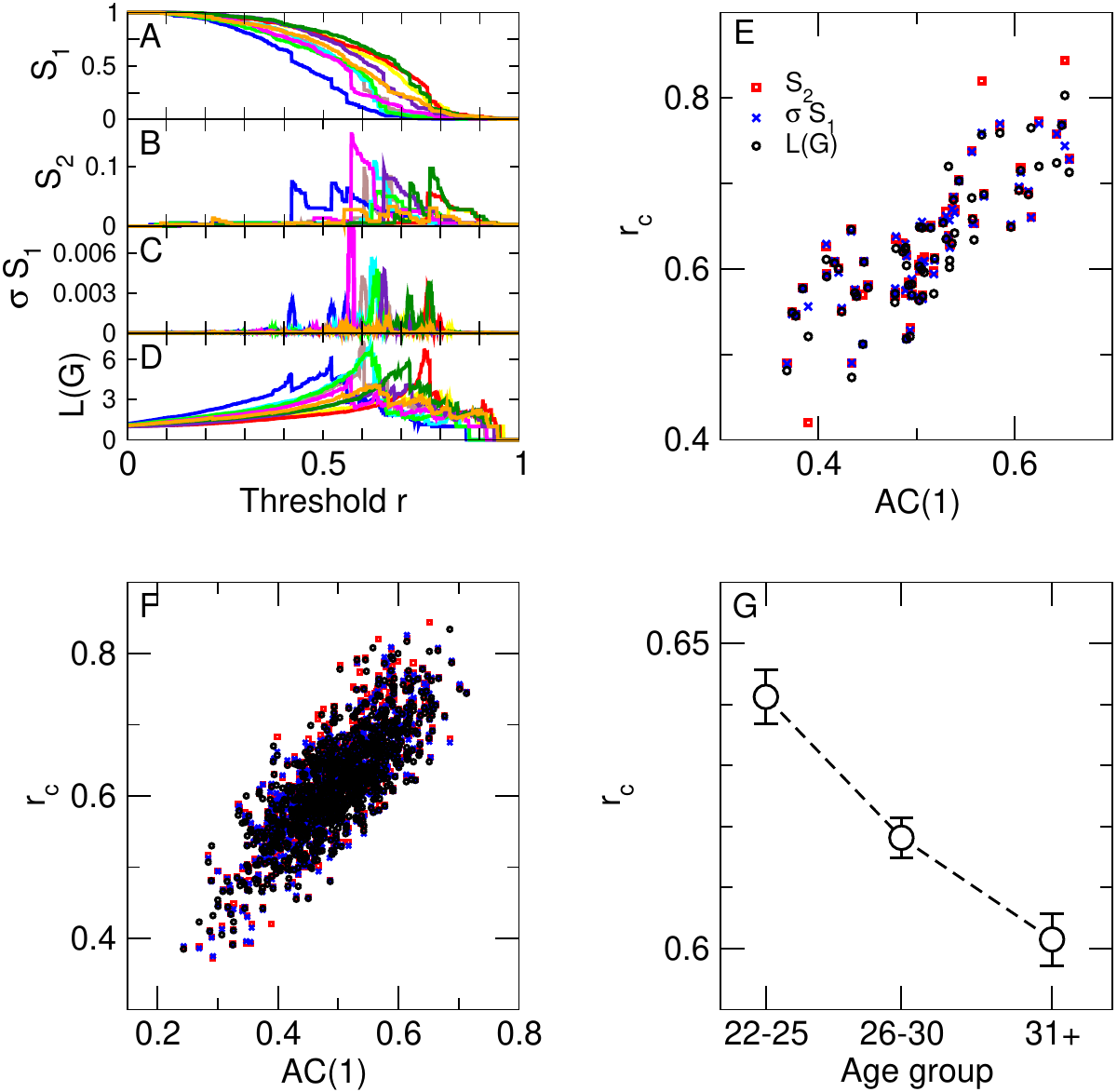}
 \caption{\textbf{Reproduction of Figures 2 and 3 from  from main text, considering concatenated brain BOLD activity of all 4 sessions for each subject in the HCP dataset}. Panels A-D reproduce Fig. 2 panels A-D, using the same color scheme for each subject (Panel A: $S_1$ as a function of $r$ for 10 subjects; Panel B: $S_2$; Panel C: $\sigma\, S_1$; Panel D: $L(G)$). Panel E reproduces Fig 2 Panel E ($r_c$ as a function of $AC(1)$ for 64 subjects, considering 3 ways to compute $r_c$). Panels F and G reproduce Fig. 3 panels A and B respectively. In Panel E, results for $r_c$  as a function of $AC(1)$ for all subjects and 3 ways to compute $r_c$ are shown.  In Panel G $r_c$ (estimated from the peak of $S_2$) is shown for 4 subjects groups, organized by age. }
    \label{figSM1}
\end{figure}

Figures 2 and 3 from main text were computed from the first session of all 996 considered subjects from WU-Minn Young Adult dataset \cite{smWuMinn}. In Fig. \ref{figSM1} we reproduce the results using all 4 sessions for each subject. To avoid introducing spurious correlations, we concatenated the timeseries of each subject after subtracting from each ROI its mean value. In each figure panel we have used the same subjects as in the related figure in main text (now considering all sessions). Results in Fig. \ref{figSM1}, panels A-F show striking similarities with their main text counterparts. Notice that in panel F we show $r_c$ estimated from $\sigma S_1$ (omitted in Fig. 3 of main text). Notice also that the results are very similar to those from $S_2$ and $L(G)$.   Results in panel G ($r_c$ as a function of age) are very similar to results in Fig 3B. In particular, the negative relation among Age group and $r_c$ is clearly observed.

\subsection{Numerical simulation results for the left hemisphere} 

\begin{figure}[ht!]
\centering
\includegraphics[width=	.4\linewidth]{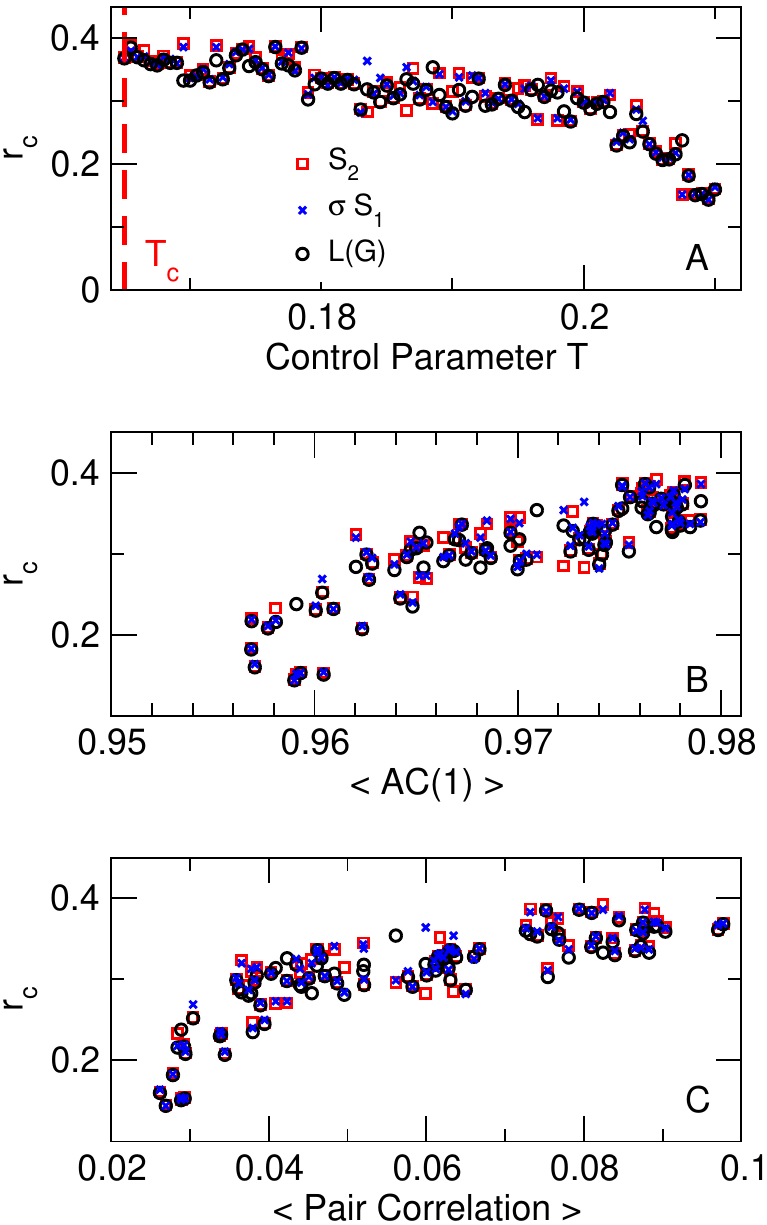}
\caption{ \textbf{GH Model simulation results from brain \emph{left} hemisphere}. Panel A: $r_c$ estimated as the peak of $S_2$, as the peak of $\sigma\, S_1$ and as the peak of the characteristic path length $L(G)$ can be successfully estimated from the percolation thresholds $r_c$. Panel B: $r_c$ estimates as a function of the first autocorrelation coefficient, $AC(1)$. Panel C: $r_c$  as a function of the average pair wise correlation among ROI timeseries. The red dashed vertical line in panel A indicates the value of the model critical point $T_c$.  Results computed  on a  representative structural network  (averaged over all subjects, as in Fig. 5 from main text). We now restricted the conectome to connections in the left hemisphere. Computation details as in Fig 5 of main text.}
 \label{figSM3}
\end{figure}

In main text Fig. 5, the numerical simulation results for the GH model \cite{smHaimovici} on the average structural connectome from WU-Minn DTI data considered only the 180 nodes of the right hemisphere. In Fig. \ref{figSM3}  we show the same results for the left hemisphere.  Computations were performed exactly as described in main text for the right hemisphere: The representative structural connectome was computed from the average of DTI connectivity matrices from the 996 subjects in the WU-Minn dataset. Then the 180 $\times$ 180 matrix corresponding to the left hemisphere (ROIs 181 to 360) was considered, and normalized as described in Rocha \emph{et al.} \cite{smrocha1,smrocha2} (i.e., the sum of the weights of in-connections of each ROI is normalized to 1). The numerical simulation of GH model was run on that network using the same parameters ($p_1=0.001$, $p_2=0.2$ $T=0.16-0.21$ with $\Delta T$ steps of $0.0005$) and duration ($10000$ steps for each $T$ value, first $1000$ steps not considered). Results in Fig. \ref{figSM3}  are very similar to those presented in  Fig. 5 of main text (for the other hemisphere).

%%%%%%%%% Figure SM3%%%%%%%%%

\newpage 
\section{Generality of the results}\label{general}
\subsection{Replication of the $r$-spectra on simple numerical models}
The inference method proposed in the main text, $r$-spectra, should be applicable in principle on any system presenting critical behavior. We show results  on two well understood numerical models.

\begin{figure}[ht!]
    \centering
    \includegraphics[width=0.6\linewidth]{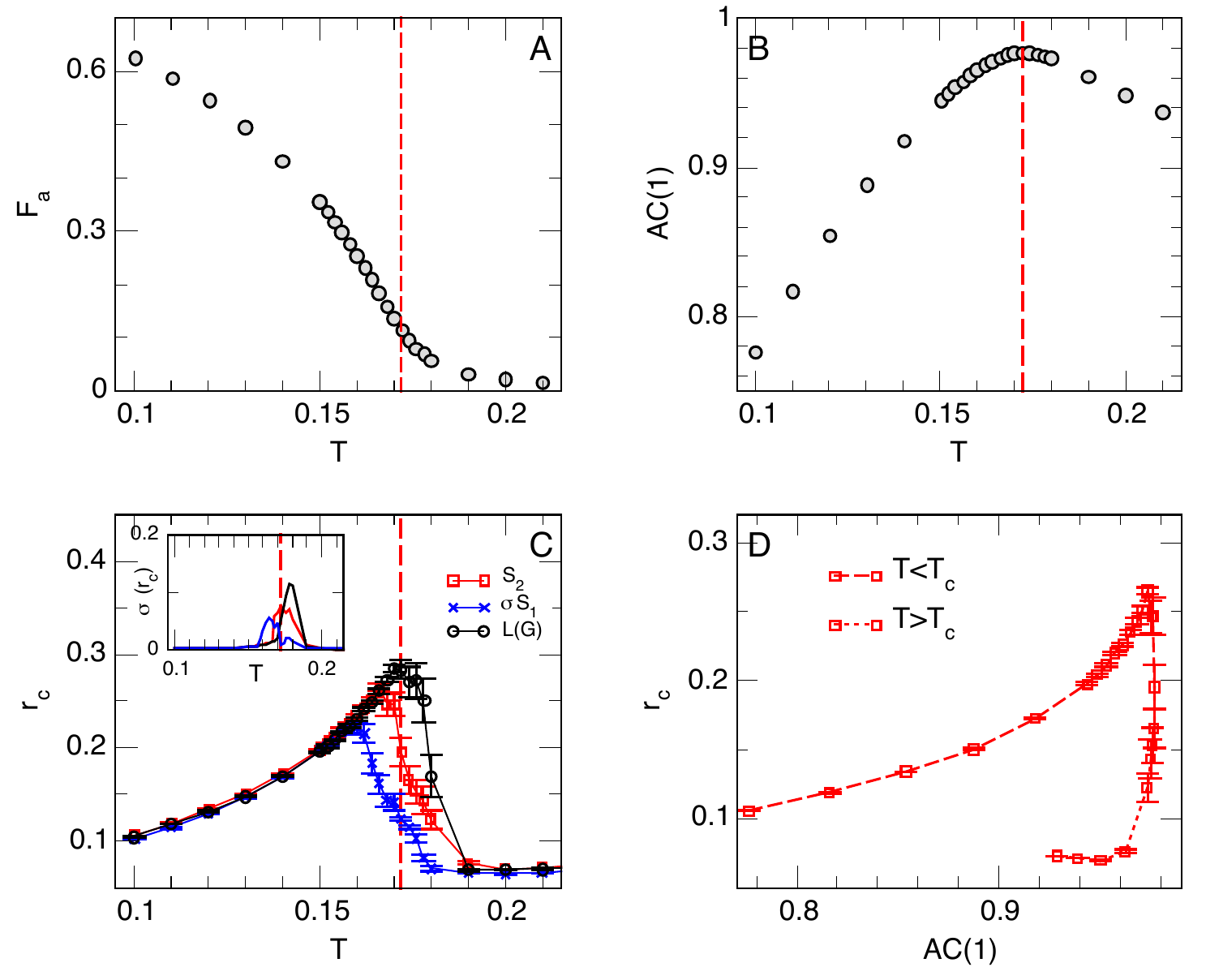}
 \caption{\textbf{Replication of the $r$-spectra framework for GH simulations on a WS network}
Panel A: Activity fraction as a function of threshold $T$. Panel B: First autocorrelation function as a function of $T$. Panel C: $r_c$ computed from three methods as a function of $T$. Inset: standard deviation $\sigma$ of $r_c$ as a function of T. Panel D: $r_c$ computed from the peak of $S_2$ as a function of $AC(1)$. In the first 3 panels, a vertical red dashed line shows the critical values of the control parameter, computed as the value that maximizes the first autocorrelation coefficient, $AC(1)$. Results  computed on a WS network of $N=5000$ nodes, with average degree $\langle k \rangle =10$ and rewiring probability $\pi=0.6$. Simulations are run for 50 000 MC steps, deleting the first $10\%$ steps. Results averaged from 10 different networks. Error bars  omitted when their size is similar to point size. }
    \label{fig:WS}
\end{figure}

We first run simulations of the GH model on a Watts-Strogatz (WS) network \cite{smWS}. While synthetic networks may not fully reproduce the properties of human connectome, they allow us to study the dynamics for arbitrary sizes. For this model and topology, a number of results showing critical dynamics at a given critical value, $T_c$, have already been described \cite{smZarepour}.

Following previous results for this model and topology, we consider networks of $N=5000$ nodes with average degree $\langle k \rangle =10$ and rewiring probability $\pi=0.6$.  The weights of connected nodes are symmetric ($w_{ij}=w_{ji}$) and are taken from an exponential distribution ($p(w_{ij}=w)\propto e^{-\lambda \omega}$, with $\lambda=12.5$).   We consider $p_1=10^{-3}$ and $p_2=0.2$ (as in main text). The model shows a continuous transition at $T_C\simeq 0.172$, from high activity  in the supercritical ($T<T_c$) regime, to low activity in the subcritical regime. $T_c$ can be identified, for instance, as the value of $T$ at which the first autocorrelation coefficient $AC(1)$ is maximum. Here, to be consistent with  results in main text, we considered that a neuron is active if it is either in the excited or quiescent state, and $p_2=0.2$. This differs from the results in previous articles for this model and topology (such as \cite{smZarepour,smSanchez}), nevertheless, this choice does not qualitatively change the results.

For each value of  $T$ in  $T=0.1-0.22$, with $\Delta T=0.01$, we run 5 realizations of  $50000$ timesteps each.
We discard the first $10\%$ time points of each timeseries and compute the Pearson correlation matrix with the remaining time points. We later compute $r_c$ for each realization as  described in main text. 

 In Fig. \ref{fig:WS} we show simulation results. In Panel A we show the activity as a function of $T$. In Panel B we show the first autocorrelation coefficient, $AC(1)$ as a function of $T$. $AC(1)$ has a peak at the transition point,  $T_c\simeq0.172$. In panel C we show $r_c$ computed in several ways, as a function of $T$. We find that the different alternative computations of $r_c$ have a peak about $T_c$, being lower at the supercritical and subcritical regimes.  We also find larger error bars compatible with higher variability about $T_c$. The inset of the figure shows $r_c$ variability among realizations, which is maximum about $r_c$. In panel D we plot $r_c$ as a function of $AC(1)$. We find that $r_c$ grows as $AC(1)$ grows, but the relation is different for subcritical ($T>T_c$) and supercritical ($T<T_c$) regimes.

\begin{figure}[ht!]
    \centering
    \includegraphics[width=0.6\linewidth]{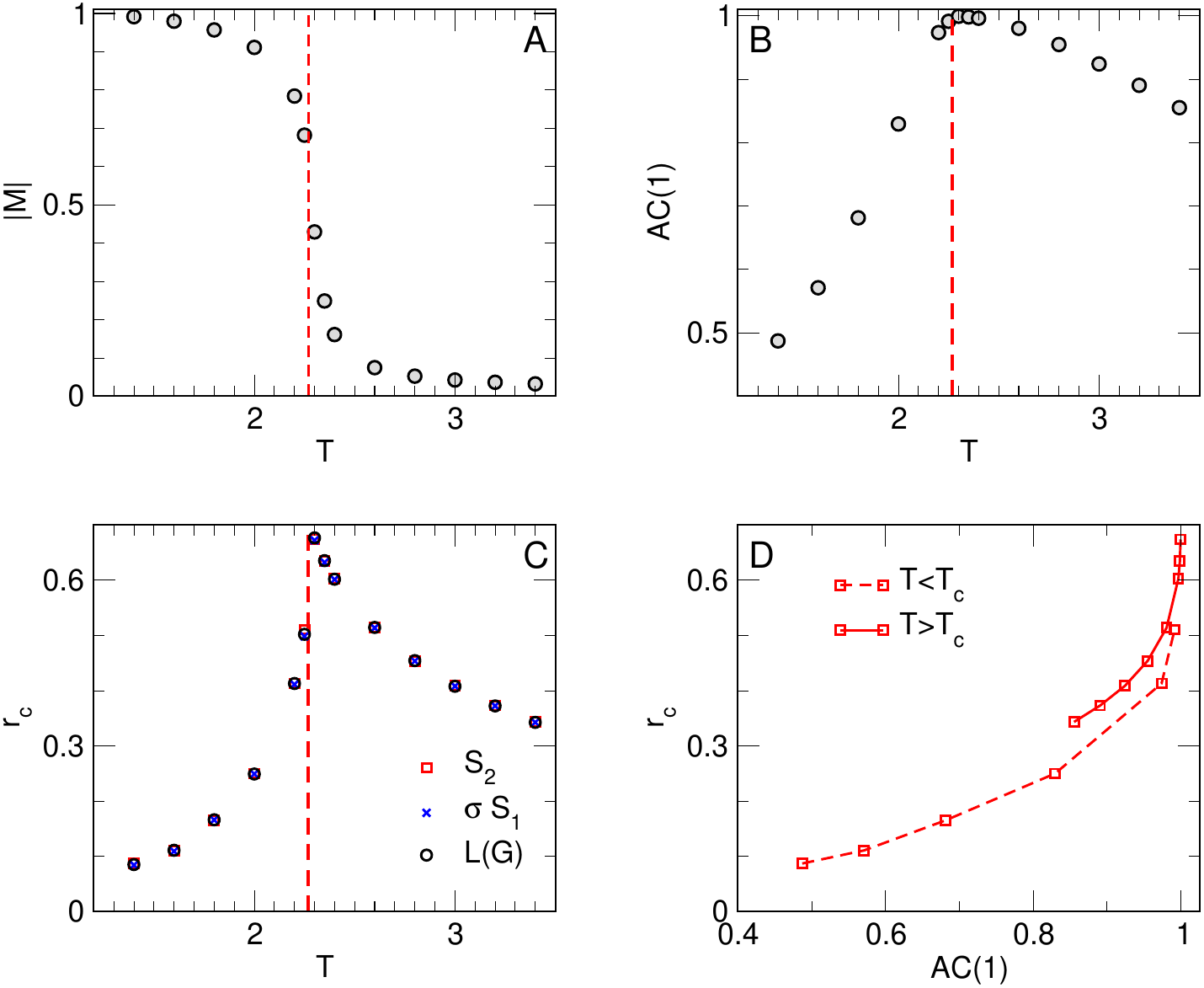}
 \caption{\textbf{Replication of the $r$-spectra framework on the 2D Ising model}
Panel A: Magnetization as a function of temperature $T$. Panel B: First autocorrelation function as a function of $T$. Panel C: $r_c$ computed from three methods as a function of $T$. Panel D: $r_c$ computed from the peak of $S_2$ as a function of $AC(1)$. In the first 3 panels, a vertical red dashed line shows the critical values of the control parameter, computed as the value that maximizes the first autocorrelation coefficient, $AC(1)$. Results  computed on a 2D lattice of $64 \times 64$ spins with periodic boundary conditions. Simulations are run for 50 000 MC steps, deleting the first $10\%$ steps. In all plots, error bars are omitted since their size is similar to point size. }
    \label{fig:Ising}
\end{figure}

We also consider the 2D Ising model, a paradigmatic example of equilibrium continuous transition. The model describes a continuous paramagnetic-ferromagnetic transition at a temperature $T_c\simeq 2.27$. Notice that we have used the same letter, $T$, for the control parameter of GH and Ising models, however, here $T$ represents temperature instead of a Threshold. We have run numerical simulations using the metropolis Monte Carlo algorithm, on a square lattice of $N=L \times L$, spins with periodic boundary conditions. For each $T$ value, we run 5 realizations of  50.000 timesteps each. To avoid transients, the first $10\%$ steps are discarded, and the remaining steps are considered for computing the magnetization $M(t)$ (the sum of the values of each spin at time $t$), the first coefficient of the autocorrelation of the magnetization timeseries, $AC(1)$, and the Pearson Correlation matrix. To reduce memory computational demands, we only save one configuration every 5 time steps (i.e., the Pearson correlation matrix is computed from $45.000/5=9.000$ configurations). This choice does not significantly change the results, while it reduces the required computer memory and also the computational time required.

For each realization and  $T$ value, and compute $r_c$ as described in main text.  In Fig. \ref{fig:Ising} we plot $r_c$ as a function of $T$ on systems of size $L=64$ (i.e., there are $N=64^2=4096$ spin timeseries). We find that $r_c$ is maximum at $T_c$, and decays both in the supercritical  ($T>T_c$) regime, and the subcritical regimes, although the relation among $AC(1)$ and $r_c$ is different for $T>T_c$ and $T<T_c$.

\subsection{Alternative definitions of $r_c$}

In main text we considered 3 ways to compute $r_c$. Nevertheless, a percolation transition can be characterized in several ways, potentially allowing alternative definitions of $r_c$. Here we show two  alternatives.

\begin{figure}[ht!]
    \centering
    \includegraphics[width=0.6\linewidth]{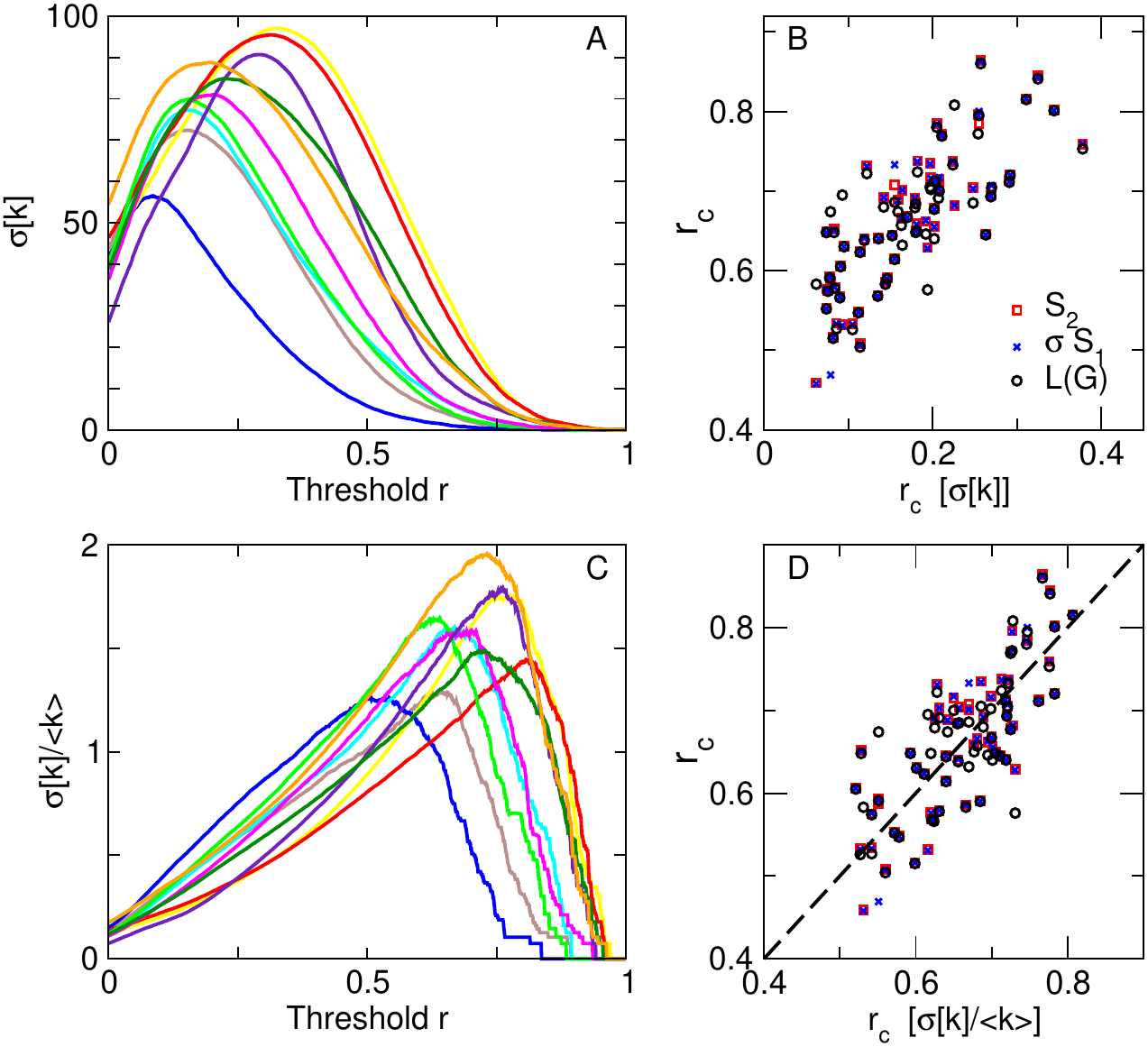}
 \caption{\textbf{Network degree variability peaks at the percolation threshold.}
Panel A: standard deviation of the degree, $\sigma[k]$  as a function of $r$. Panel B: $r_c$ as a function of $r_c$ estimated from the peak of $\sigma[k]$. Panels C and D: same as A and B for $\sigma[k]/\langle k\rangle$. Data from the first session of the  HCP dataset, considering the same  10 (panels A and C) and 64 (panels B and D) subjects and the same colors for each one, as in Figs. 2 and 3 on main text.}
    \label{fig:sigmak}
\end{figure}

We first consider the degree (i.e., the number of links of each node) distribution, $\{k_i \}_{i=1,..,N}$ in the graph obtained from the binarized matrix.  Close to the percolation point, a broad distribution of degrees is expected, and consequently a large variability in the degree should be present. In Fig. \ref{fig:sigmak} we show the standard deviation of the degree as a function of $r$ from the correlation matrix of 10 subjects in the HCP dataset. We observe that the curves are smooth, and show a broad peak. We propose this peak as an alternative definition of $r_c$. In Panel B we observe that this definition correlates well with the already proposed definitions, although its value is smaller.

We may also define $r_c$ as the peak of  coefficient of variation: the ratio among the standard deviation of the degree, $\sigma[k]$ and its mean, $\langle k\rangle$ (see Panels C and D). This definition is still smoother than those used in main text, and shows  quantitative values similar to those considered in main text. Notice that both definitions, computed from degree distribution, do not require cluster computation and are consequently numerically more efficient.

\begin{figure}[ht!]
    \centering
    \includegraphics[width=0.6\linewidth]{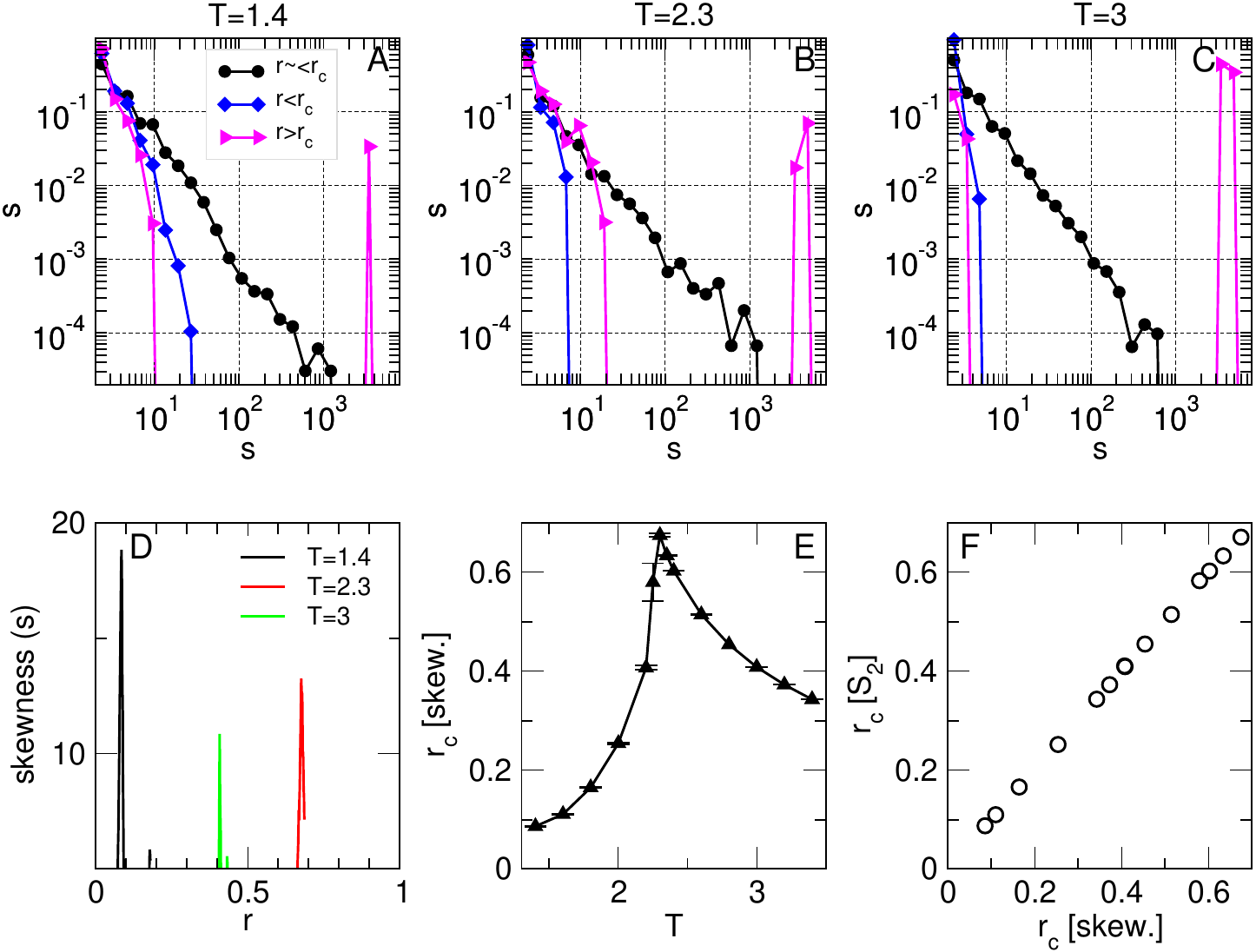}
 \caption{\textbf{Alternative estimation of $r_c$ from the cluster size skewness, demonstrated in the  Ising model.}
Panel A: Cluster size distribution $P(s)$ from the percolation of the correlation matrix of Ising model dynamics for $T=1.4$, for 3 values of $r$. Panels B and C: same as Panel A for $T=2.3$ and $T=3$. Panel D: Skewness of $P(s)$ for the 3 temperatures, as a function of $r$. Panel E: $r_c$ estimated as the peak of the Skewness, as a function of $T$. Panel F: $r_c$  computed from $S_2$ as a function of $r_c$ computed from the skewness of $P(s)$. Each point represents the average over 5 realizations of the same $T$ value.  In panels A-C, to increase statistics, for each $r$ value, we average over all cluster sizes computed from $r-0.005$  to $r+0.005$. Further numerical details as in Fig.\ref{fig:Ising}.}
    \label{fig:IsingSkew}
\end{figure}

Another possible definition of $r_c$ is related to cluster size distribution. At the percolation point, a scale-free distribution of cluster sizes is expected, while fast decaying distributions are expected for subcritical regimes, and giant clusters, including a large proportion of nodes, are expected for the supercritical regime. As a consequence, the skewness of the cluster size distribution should be maximum near the percolation point. Using this property we define an alternative definition of $r_c$: the maximum of the skewness of the cluster size distribution as a function of $r$. Importantly,  a large number of ROIs is required for this definition to be useful  (since a reliable cluster size distribution needs to be computed). Also, clusters of size 1 (i.e., isolated ROIs) are excluded. Results for Ising model are shown in Fig. \ref{fig:IsingSkew}. We find that the values of $r_c$ computed with this definition  coincide with the estimate from the peaks of $S_2$ (see panel F).

\newpage
\section{Percolation results}
\label{percolation}

\subsection{Relation to standard percolation} 
In main text we focused on networks derived from the thresholding of the  correlation matrix, which showed behavior compatible with a percolation transition as the threshold value $r$ is varied. 
Percolation phenomena is most frequently studied  as a function of  the density of preserved links $p$.  The relation among $p$ and $r$ is studied next.

\begin{figure}[ht!]
    \centering
    \includegraphics[width=0.6\linewidth]{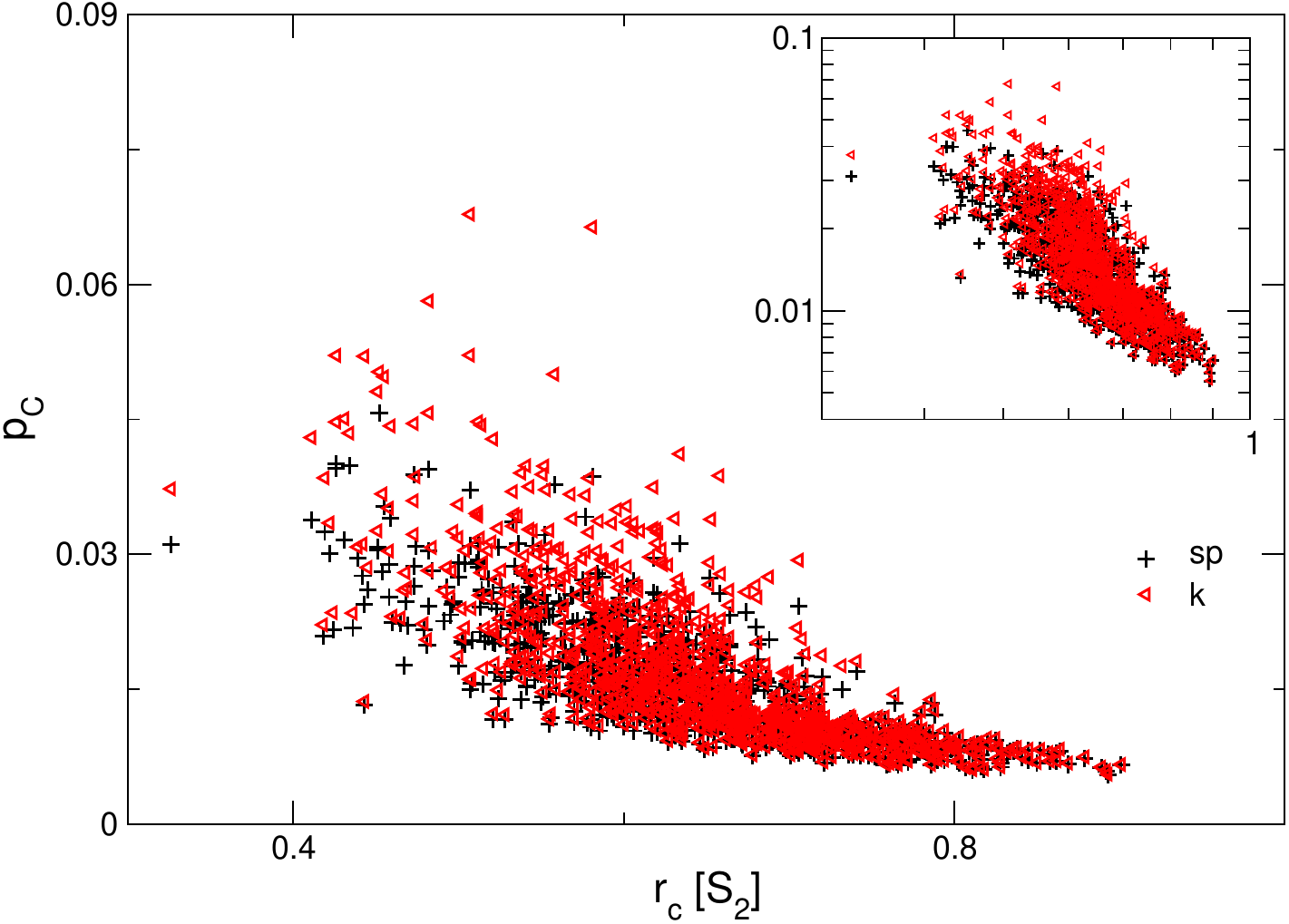}
 \caption{\textbf{Relation among $r_c$ and $p_c$ for HCP subjects.} $p_c^{(sp)}$ and $p_c^{(k)}$ computed from the correlation matrix of each subject, as a function of $r_c$. The inset shows the same data in logarithmic scale. The first session of all 996 HCP  subjects was considered.}
    \label{fig:pcrc}
\end{figure}

Analytical approximations predicting the critical $p_c$ value from an adjacency matrix $A$ are known \cite{smRadicci}. One of such approximations is $p_c^{(sp)}=1/\lambda_1$, being $\lambda_1$ the largest eigenvalue of $A$. Another approximation is given by $p_c^{(k)}=\frac{\langle k\rangle}{\langle k^2\rangle-\langle k\rangle}$, where $k_i$ is the degree of the $i-th$ node of the (unweighted) network. We computed $p_c^{(sp)}$ using the correlation matrix instead of $A$, and $p_c^{(k)}$ replacing $k_i$ with the strength of each node, $S_i=\sum_j C_{ij}$. These estimated critical densities show a clear (inverse) correlation with the respective $r_c$, suggesting a connection among the proposed method and standard percolation on graphs, see Fig. \ref{fig:pcrc}.

\subsection{Cluster size variability}

In this section we show further evidence on the existence of a percolation transition as  the threshold of correlation matrix, $r$, is varied.  We compute the variability among subjects of the largest cluster size and the number of clusters from the r-spectrum of the subjects in the HCP dataset.

\begin{figure}[ht!]
    \centering
    \includegraphics[width=0.4\linewidth]{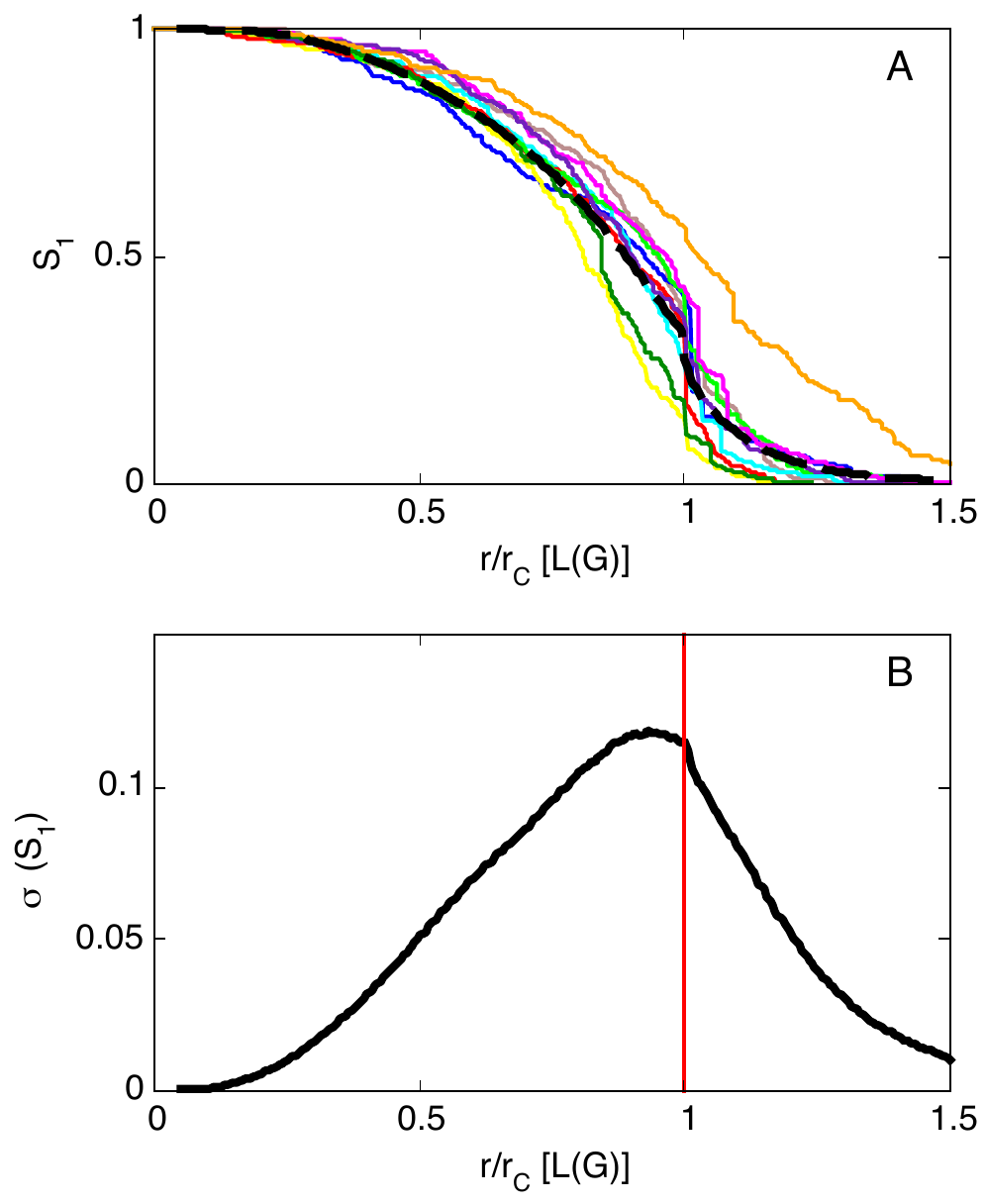}
 \caption{\textbf{Largest cluster size variability at percolation.}
Panel A: Largest cluster size, $S_1$ as a function of $r/r_c$ for 10 representative subjects (colors). A dashed black line shows the average over all 996 subjects.  Panel B: Standard deviation of $S_1$ among subjects, for the same value of $r/r_c$. A red vertical line shows $r=r_c$.  Results  computed from the percolation of the correlation matrix of the fMRI activity of the first session of the HCP dataset. The same 10 representative subjects of Fig. 2 were used in panel A, following the same color scheme. }
    \label{fig:Variab}
\end{figure}

In Fig. \ref{fig:Variab}-A we show $S_1$ as a function of $r/r_c$ for 10 representative subjects (the same subjects and color code as in Fig. 2). For each subject we used the $r_c$ estimate from $L(G)$.  
We observe that, for 9 out of the 10 subjects, $S_1$ as a function of $r/r_c$ seems to collapse.  
We have tried different alternative definitions of $r_c$ (such as $r_c$ from the peak of $S_2$ or $\sigma S_1$). None of them can collapse the $S_1$ curve for all subjects. We have also investigated ways to reduce sources of variability in the data. We were unable to improve the collapse. These results suggest that the differences in the $S_1$ v.s. $r$ curves may be due to true dynamical differences among subjects.

In panel B we show the among-subject variability of $S_1$ and the number of clusters as a function of $r/r_c$. We observe that both curves present a peak about $r=r_c$.  Similar results hold for the number of clusters and its variability (provided we exclude clusters of size 1, which represent isolated ROIs).

\section{Relation between Pearson correlation and Time autocorrelation.} \label{correlations}

Higher time autocorrelation is expected for the sum of signals that show high Pearson correlation among them. This relation between Pearson correlation and autocorrelation is predicted, for instance, in the context of continuous phase transitions, where dynamic scaling is assumed.

Analytical results can be found in simpler cases, such as VAR(1) models studied in \cite{smrefAC_CC}.

Consider $n$ timeseries $x_i(t)$, where $i=1,..,n$, whose dynamics is given by 

\begin{equation}
    x_i(t)=\sum_j A_{ij} x_j(t-1) + \xi_{i,t}, \label{VAR1}
\end{equation}
where $\xi_i$ are zero mean, uncorrelated random variables, and it is assumed that the module of the eigenvalues of $A$ are smaller than one. It can be found that \cite{smrefAC_CC}

$$\mathbf{C}(1)=\mathbf{C} \times \mathbf{A}^T,$$
where $C_{ij}=\langle x_i(t) x_j(t) \rangle$, $C_{ij}(1)=\langle x_i(t) x_j(t+1) \rangle$  and $\langle ... \rangle$ stands for time average. The first autocorrelation coefficient can be written as

$$AC(1)=\frac{\sum_{ij} C_{ij}(1) -\langle S\rangle^2 }{var(S)},$$
where $S(t)=\sum_i x_i(t)$. Similarly, the Pearson correlation matrix $\mathbf{P}$ has elements

$$P_{ij}=\frac{C_{ij}-\langle x_i\rangle \langle x_j\rangle}{\sigma_i \sigma_j}$$

Multiplying the former equation by the transpose of $\mathbf{A}$, assuming that all signals have a similar mean $\langle x_i\rangle \simeq m$ and standard deviation, $\sigma_i\simeq \sigma$, and that $var(S)\simeq n\sigma^2$, we find that

\begin{equation}
    AC(1) \simeq \frac{1}{n}\sum_{i,j} [\mathbf{P} \times \mathbf{A}^T]_{ij}. \label{RelAC_Pear}
\end{equation} Numerical simulations of Eq. \ref{VAR1} show that Eq. \ref{RelAC_Pear} holds, and also, that for positive average matrices $A$, simulations with higher Pearson correlation show higher $AC(1)$.

\section{Cluster localization near the percolation threshold} \label{localizaition}

\begin{figure}[ht]
    \centering
\includegraphics[width=0.4\linewidth]{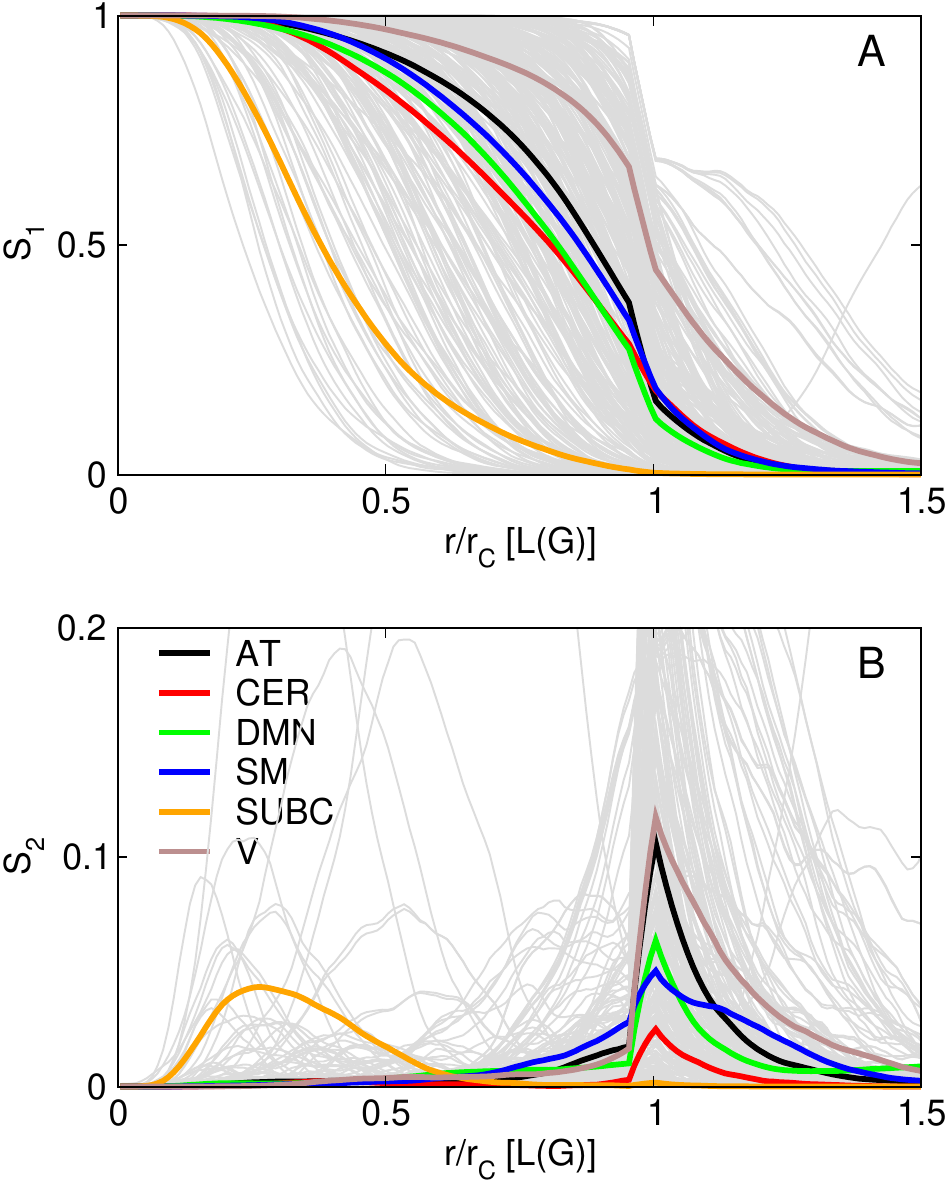}
        \label{fig:suRSN}
     \caption{\textbf{RSN participation probability as a function of $r$. } 
Fraction of nodes from each RSN belonging to the first (Panel A) or second (Panel B) largest cluster, averaged over all 996 HCP subjects. Gray lines show the same results for each ROI. All details as in previous figures. The number of ROIs per RSN are: AT (Attention) 81, CER (Cerebellum) 16, DMN (Default mode network) 91, SM (Somatomotor) 95, SUBC (Subcortical) 18, and V (Visual) 58.}
\end{figure}

Since $r_c$ estimates the percolation threshold it is expected that at that $r$ value the brain must exhibit the largest fluctuations (and dynamical flexibility). To evaluate which brain regions fluctuate in and out of the largest  clusters (i.e., $S_1$ and $S_2$) we compute the probability to belong to each cluster for each resting-state network as a function of $r$. Results in Fig. \ref{fig:suRSN} show the hierarchy in which RSNs participate (i.e., have a large number of ROIs leaving the largest cluster near the value of $r_c$). Notice that while all cortical RSNs fluctuate near $r_c$,  subcortical regions only exhibit changes for very low values of $r$.

\end{document}